# Hydrogen at GaN(0001) surface control of Fermi level pinning: Mg activation of p-type conductivity – Nakamura process deciphered

Konrad Sakowski, Pawel Strak, Pawel Kempisty, Izabella Grzegory, Stanislaw Krukowski

Institute of High Pressure Physics, Polish Academy of Sciences, Sokołowska 29/37, 01-142 Warsaw, Poland

**Abstract**

*Ab initio* calculations were used to disentangle the mystery of Nakamura activation of p-type in Mg doped MOVPE grown gallium nitride, the key process leading to the 2014 Nobel Prize in Physics. Calculations were used to obtain the equilibrium state of the hydrogen atom deep in the GaN bulk and at the GaN(0001) surface. It was shown that the H position within bulk GaN depends on the Fermi level: in n-type GaN, it is located in the channel, whereas in p-type GaN, it is attached to the N atom, breaking one of the GaN bonds. In contrast, at the GaN(0001) surface, H is attached in the on-top position for any hydrogen coverage; for low and high H-coverage, the Fermi level is pinned at the Ga – broken bond state and at the valence band maximum (VBM), respectively. The diffusion path from the bulk to the surface was obtained when the Fermi level was high and low, the barrier was zero, and $\Delta E_{bar} \approx 1.717\ eV$, which effectively blocked hydrogen escape into the vapor. Thus, high H coverage, that is, high hydrogen pressure in the vapor, prevents H from escaping from the bulk to the surface, whereas at low coverage (low hydrogen pressure), the process is barrierless. It is therefore proven that the hydrogen escape control step in the Nakamura process is the transition of hydrogen from the bulk to the surface, which is controlled by the position of the Fermi level at the surface. Molecular hydrogen desorption from the surface is easy for high H coverage and difficult for low, thus opposite to observed experimentally thus this process is not the determining step in activation. A full thermodynamic estimate of the maximal partial pressure of hydrogen in the vapor, corresponding to the transition of the Fermi level from the Ga-broken bond state to the VBM, was used to establish the maximal hydrogen pressure limit for the p-type Mg activation process.

## I. Introduction

An important barrier to the technological success of bipolar nitride-based devices is related to the hole conductivity of group III nitrides. The absence of p-type doping has blocked the development of nitrides for decades [1,2]. The first breakthrough occurred when Amano and Akasaki discovered the activation of p-type by electron irradiation of Mg-doped MOVPE GaN grown on sapphire [3]. This was followed by the discovery of hole conductivity activation via hydrogen-free annealing of Mg-doped nitride structures by Nakamura [4-6]. This led to the extensive development of nitride light sources that were honored by the 2014 Nobel Prize awarded to H. Amano, I. Akasaki, and S. Nakamura after a considerable period of time [7]. Nevertheless, much more time passed, and the microscopic explanation of the Nakamura processes was not yielded.

The successful incorporation of magnesium in the metal organic vapor phase epitaxy (MOVPE) growth of GaN was explained by the passivation of Mg acceptors by the creation of electrically neutral complexes with hydrogen (Mg-H) during growth [8]. The activation of the p-type occurs via the physical removal of hydrogen from the solid into vapor [9]. The process includes several steps: (i) dissociation of the Mg-H complex, (ii) diffusion in the GaN bulk, and (iii) surface desorption of hydrogen ($H_2$) di-molecules. Combined *ab initio* and Fourier Transform Infrared Spectroscopy (FTIR) studies showed that the Mg-H complex and separate $H^+$ ions dominate at low and high temperatures, respectively [10]. The Secondary Ion Mass Spectrometry and Hall mobility measurements indicate that Mg-H dissociation barrier is relatively high ($\Delta E_{bar}(Mg - H) \approx 0.8 \div 1.5\ eV$) for low temperatures and low ($\Delta E_{bar-dis}(Mg - H) \approx 0.2 \div 0.5\ eV$) for high temperatures. The transition temperature is ($T \approx 600°C$) which allows for effective activation just above this temperature [11]. It was found that the diffusion barrier of the separate ion $H^+$ is relatively small, of the order $\left(\Delta E_{bar-dif}(H^+) \approx 0.4\ eV\right)$ [12]. Because the sum of the ion $H^+$ diffusion barrier and binding energy to the occupied $Mg^-$ acceptor is $1.76\ eV$, the latter is $\Delta E_{bar-act}(H^+) = 1.36\ eV$. In addition, it was concluded that the H-coverage of the GaN(0001) surface has to be relatively low for the efficient activation of p-type GaN [9]. It was also concluded that the energy of the adsorbed H adatom at the GaN(0001) surface is close to that of the molecular $H_2$ hydrogen in the vapor [9]. Thus, a detailed study of these processes led to the conclusion that the surface reaction is most likely the rate-controlling step [13]. These two factors can control the process: surface potential barrier, transition to the surface state, and creation of a di-molecule. The potential barrier was measured using photoelectron spectroscopy (PES) for both n-type and p-type GaN grown by MOVPE [14]. The Fermi level was located 2.55 $eV$ above the valence band maximum (VBM). The results were identical for all conductivity types, which indicates that the Fermi level is pinned by the surface state located 0.92 $eV$ below the conduction band minimum (CBM). In the case of n-type materials, the bulk position of the Fermi level is located at about 0.02 $eV$ below the CBM [15]; thus, the upward band bending at the surface could be estimated to $\phi \approx 0.90\ eV$. In the case of p-type GaN obtained by Mg doping, the position of Fermi level is about 0.18 $eV$ [16] so that the downward band is estimated as $\phi \approx 2.37\ eV$. In both cases, it was considerable, nevertheless in Ref. 13 it was concluded that this is a relatively minor effect, so that the overall activation rate is controlled by the final step, that is, the creation of di-molecules at the surface [13]. In addition, it was determined that this was a second-order reaction in the concentration of hydrogen, clearly indicating that the control was due to the creation of di-molecules.

P-type doping by Mg is far from perfect. Owing to its high activation energy equal to 0.18 $eV$ about one percent is occupied at room temperature [16]. In addition, it is possible that the decomposition of Mg-H complexes is not complete [17]. This could be retarded by blocking hydrogen diffusion in n-type layers [18]. Finally, compensation by non-intentionally incorporated donors may be significant [19].

In this study, we used ab initio calculations to determine the influence of the surface potential barrier and the creation of hydrogen di-molecules on the removal of hydrogen from GaN. These data are based on the analysis of the properties of the GaN(0001) surface under various H coverages. Using these calculations and thermodynamic analyses, the conditions for hydrogen necessary for effective p-type activation by annealing were derived. The nature of the Nakamura process of p-type activation in Mg-doped MOVPE-grown GaN will be elucidated.

## II. The calculation procedure

*Ab initio* density functional theory (DFT) calculations were applied to large GaN slabs to simulate the adsorption of ammonia and hydrogen at the GaN(0001) surface. The Spanish Initiative for Electronic Simulations with Thousands of Atoms (SIESTA) *ab initio* code is used to solve the Kohn-Sham set of equations [20]. The eigenfunctions were obtained as linear combinations of finite-size numeric atomic orbitals [21, 22]. The angular dependence is represented by spherical harmonics, i.e. *s, p* and *d* orbitals. The *s* and *p* orbitals of the gallium and nitrogen atoms are represented by triple zeta functions. In the case of gallium, the internal $d$ shell electrons are used as the valence electrons, which are represented by single zeta functions. The functional basis is small owing to the application of Troullier-Martins pseudopotentials [23,24]. The integration in k-space is approximated by a sum over a Monkhorst-Pack grid (1 × 1 × 1) [25]. This reduction is essentially an application of the 19[th] century Gauss quadrature [26]. The GGA-PBE (PBEJsJrLO) functional was parameterized using the β, μ, and κ values set by the jellium surface (Js), jellium response (Jr), and Lieb-Oxford bound (LO) criteria [27,28]. In the calculation, the SCF loop was terminated when the difference for all elements of the density matrix in two consecutive iterations wais smaller than $10^{-4}$. A real-space grid is used for the calculation of the multicenter overlap integrals that are controlled by the energy cut-off. The cutoff value is 410 Ry, which is equivalent to a grid spacing in the real space of 0.08 Å.

The electric potential problem, i.e. the Poisson linear equation, is solved using a Fast Fourier Transform (FFT) series, as it is far the best for linear algebra. This implies periodic boundary conditions that create nonlocal coupling across the vacuum space owing to the surface slab dipole-induced potential difference [29]. The periodicity induces interaction between slab copies that were zeroed either by introducing a compensating dipole in the vacuum space [29,30] or by the additional Laplace solution contribution to remove the field in the space between the copies [31].

The Born-Oppenheimer approximation was enforced in the procedure for determining the energy barriers. The method is based on the nudged elastic band (NEB) method to optimize the path with fixed endpoints fixed [32-34]. The minimum energy pathways (MEP) are characterized by at least one saddle point, finding the energy barrier equivalent to the notion of the activated energy complex in diffusion theory. The NEB module was linked to SIESTA to determine the minimal barrier energy and path of the species. Atom relaxation terminates when the forces acting on the atoms are not higher than 0.005 eV/Å.

The *ab initio* GaN lattice parameters were $a = 3.21$ Å and $c = 5.23$ Å being in good agreement with the experimental values: a = 3.189 Å, c = 5.186 Å [35]. The band correction scheme of Ferreira et al., known as GGA-1/2, was used to obtain the proper band gap energies, effective masses, and band structures [36,37]. The *ab initio* bandgap was $E_g^{DFT}(GaN) = 3.47\ eV$, in agreement with the low-temperature experiment $E_g^{DFT}(GaN) = 3.47\ eV$ [38,39]. These electronic properties were obtained using a modified Ferreira's scheme, for which the positions of atoms and a periodic cell were first obtained using the PBEJsJrLO exchange-correlation functional.

### III. The results.

#### a. Hydrogen in the bulk

The properties of hydrogen impurities inside the GaN lattice have been intensively investigated in the past [8-18,40-44]. In a defect-free GaN lattice, hydrogen was found to be predominantly in a single-atom configuration with the location and charge depending on the position of the Fermi level [42,43]. In Ref 42, the authors found the hydrogen in two charge states to be stable: for p-type, it is located close to the N atom, not associated with any electrons ($H^+$), in the N-antibonding site; for n-type, it is located in the Ga antibonding site, with the two occupied quantum states ($H^-$) [42]. This state is associated with a very high migration barrier $\Delta E_H^{bar} = 3.4\ eV$, preventing hydrogen diffusion [42]. In addition, it was found that the hydrogen atom in the neutral state, which is relatively high in energy and therefore rarely attained, is located in the channel. In Ref. 43, neutral H atoms have been confirmed to have high energies [43]. In p-type GaN, the positive ($H^+$) ion was found to be equally stable at bond-centered and antibonding sites, breaking the Ga-N bond [43]. In n-type GaN, the double-occupied ($H^-$) ions sit far away from the nitrogen atoms [43]. Thus, hydrogen impurities are sometimes referred to as amphoteric defects with a negative Hubbard U term [41, 44]. Additionally, the hydrogen $H_2$ molecule was found to have a much higher energy and is not stable in wurtzite GaN, and thus is not relevant for GaN properties [42,43].

In the present calculations, the energetically stable sites of hydrogen atoms located deep in the GaN bulk (under two double atomic layers (DALs) below the GaN(0001) surface) were investigated. These results were obtained for $(2\sqrt{3} \times 2\sqrt{3})$ 8 double Ga-N atomic layer (DALs) thick slab terminated by a hydrogen fractional charge ($Z = 3/4$) pseudoatoms. Thus, this state was considered to be essentially representative of the hydrogen atoms in the bulk. Thus, the slab represents a Ga-terminated GaN (0001) surface with the possibility of simulating the transition from the bulk to the surface. The relatively small thickness of the slab did not allow us to simulate surface-band bending [45]. Thus, the position of the Fermi level was determined by the hydrogen coverage of the (0001) GaN surface. Two variants of hydrogen coverage were used: (i) clean and (ii) 11 hydrogen adatoms, located on top of the Ga surface atoms. Accordingly to earlier obtained results, Fermi level at clean GaN(0001) surface is pinned by Ga broken bond state, of the energy about 0.4 eV below conduction band minimum (CBM) [45,46]. Thus, this case corresponds to n-type GaN. In contrast, the 12-sites slab with 11 attached hydrogen atoms corresponds to a high hydrogen coverage ($\theta_H = 11/12$). Thus, the Fermi level is pinned by the hydrogen state arising from the H adatom bonding to the topmost Ga surface state, with energy located at the valence band maximum (VBM) [45-47]. Thus, the Fermi level is located at the VBM, i.e. it corresponds to p-type GaN.

The energetically stable position of a hydrogen atom located under two double atomic layers of an n-type slab (clean GaN(0001) surface) is shown in Fig. 1. As can be seen, the hydrogen atom was located in the channel without any particular bonding to any other specific atom. Therefore, the bonding is mostly related to the presence of charge within the entire solid body. Thus, the results are consistent with those of Wright in Ref. 43 [43].

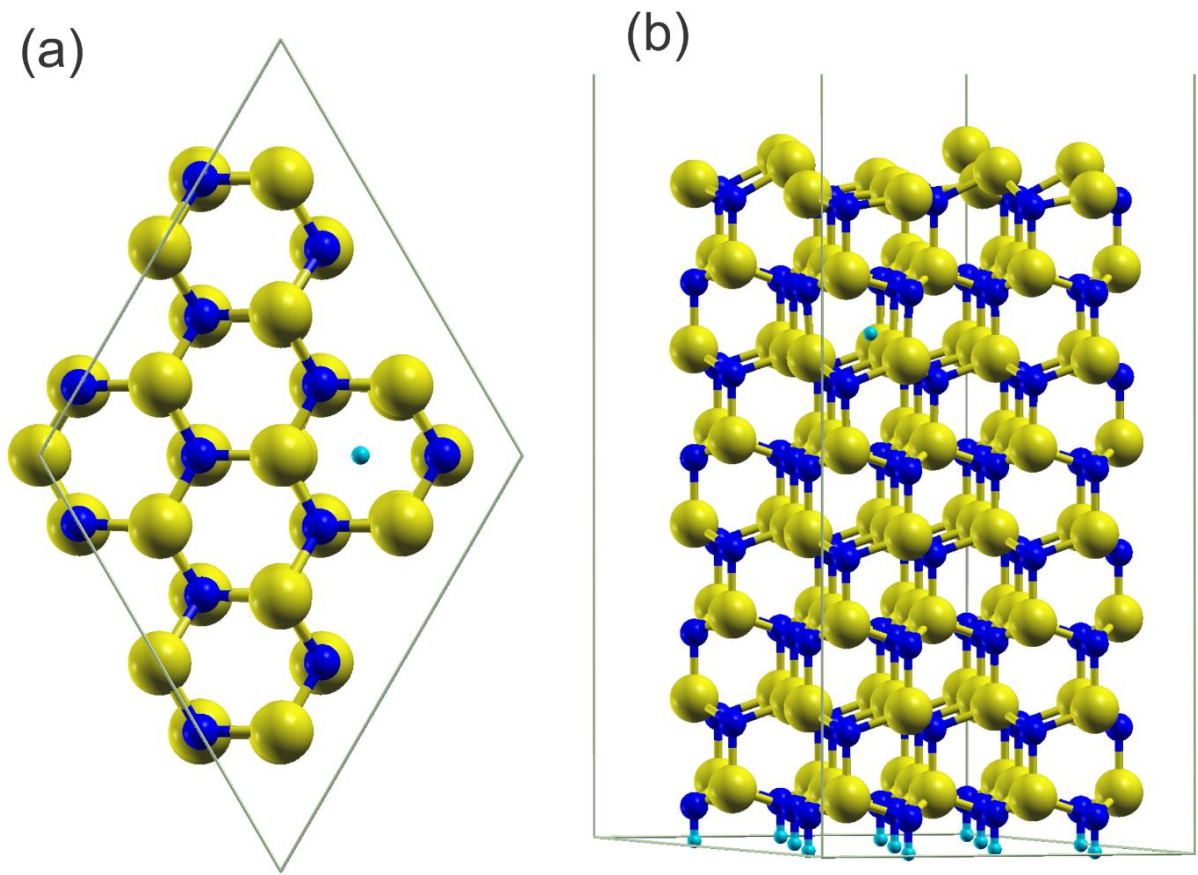

*Fig. 1. $(2\sqrt{3} \times 2\sqrt{3})$ 8 double Ga-N atom layer slab, representing clean GaN(0001) surface: (a) top view, (b) side view. Additional hydrogen atom is located under two double atomic layers (DALs). The opposite surface is terminated by fractional charge $(Z = 3/4)$ atoms. Gallium and nitrogen atoms are represented by yellow and blue balls, hydrogen atoms are represented by cyan balls.*

A $(2\sqrt{3} \times 2\sqrt{3})$ 8 double Ga-N atom layer slab was used to obtain the principal physical properties of hydrogen via *ab initio* spin-polarization calculations. The simulation results are shown in Fig. 2. As it is shown, the two spin hydrogen states have the same energy, located in the bandgap, about 1.5 eV below conduction band minimum (CBM). Because the Fermi level is pinned by Ga broken bond states of energy approximately 1 eV higher, these two states are occupied. The slab is electrically neutral with no electric field inside; therefore, the bands are flat. There are no other states associated with hydrogen atoms; therefore, these two states are the only ones that are associated with the impurity. It has to be added that hydrogen states are not localized without any noticeable potential disturbance inside the slab, and the role of the charge is merely the screening of the H nucleus, i.e. positively charged proton.

*Ab initio* calculations also revealed the second energy minimum of the hydrogen atom located close to N atom, with its energy 0.27 eV higher, therefore the second one is not minimal energy location. Nevertheless, this position is visible in Fig. 3 because this second site is the minimal energy position for p-type GaN, as discussed below.

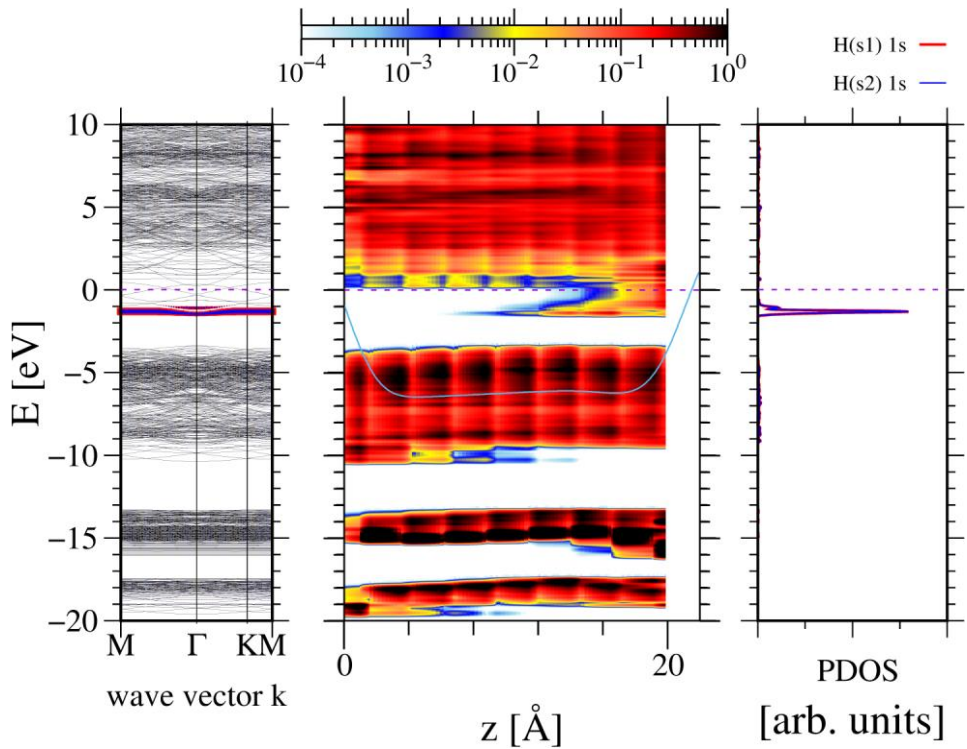

*Fig.2. Band diagram* $(2\sqrt{3} \times 2\sqrt{3})$ *8 double Ga-N atom layer slab, obtained in full spin ab initio calculations, representing clean (0001) GaN surface: left – in momentum space, middle – in real space, plotted along c-axis, surface is at the right edge of the diagram, right – projected*

*density of states (PDOS) on hydrogen two different spin value hydrogen 1s states. The scale of colors at the top represents the density of states in real space.*

A different picture is obtained for the p-type slab, i.e. 12 sites Ga-N slab covered by 11 H adatoms, as presented in Fig. 3. As shown, the hydrogen atom is located close to the N atom in the antibonding site and is bound to this atom. This bond causes disintegration of one gallium-nitrogen bond so that the Ga-N states are replaced by N-H states. These results are in agreement with the results obtained in previous studies for p-type GaN [42,43].

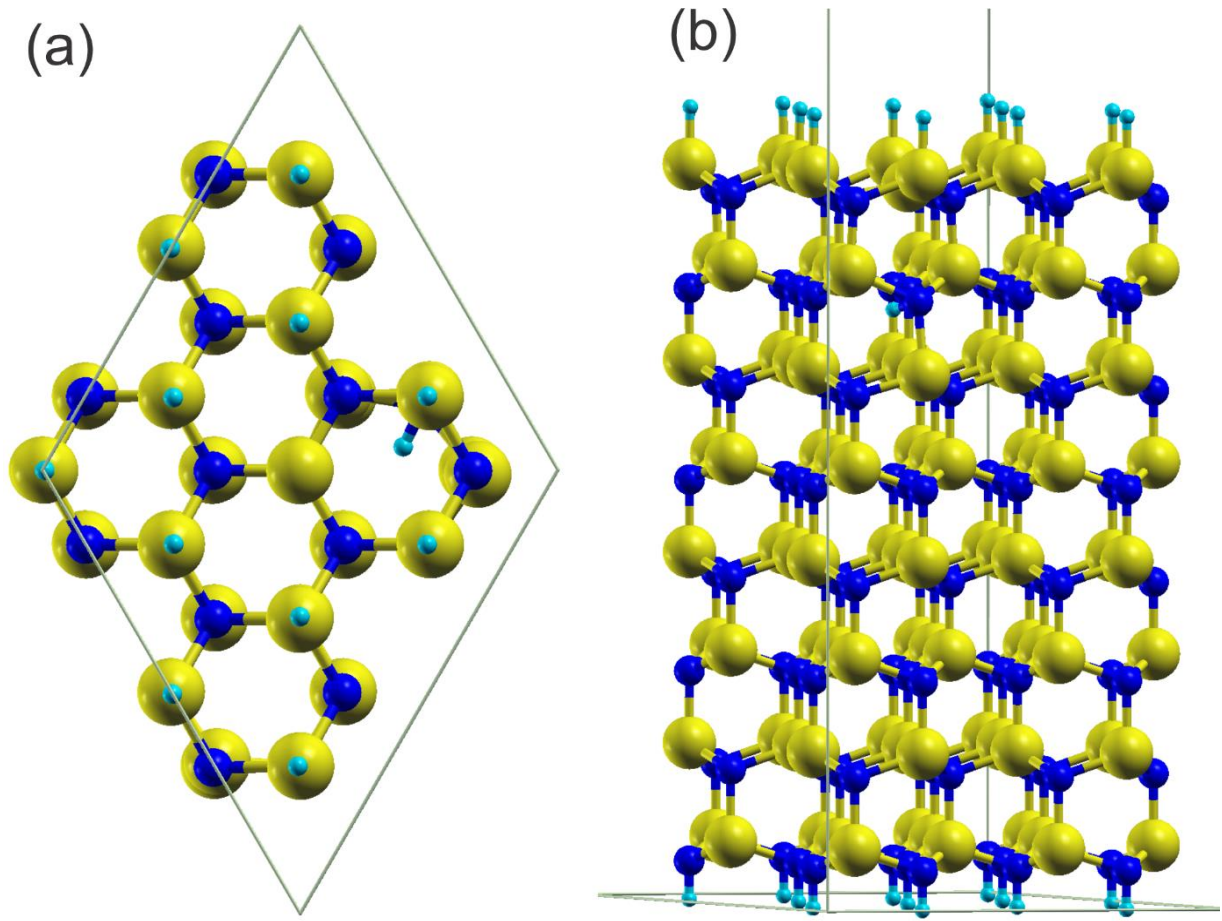

*Fig. 3. $(2\sqrt{3} \times 2\sqrt{3})$ 8 double Ga-N atom layer slab, representing 11 H atom covered GaN(0001) surface: (a) top view, (b) side view. Additional hydrogen atom is located under two double atomic layers (DALs). The opposite surface is terminated by fractional charge $(Z = 3/4)$ atoms.*

The electronic properties of the system composed of a GaN slab with H impurities obtained by spin-polarized *ab initio* calculations are presented in Fig. 4. As expected, the Fermi level is pinned by 11 H adatom states located at the VBM [45-47]. Thus, the system represents a p-type GaN. The results show drastically different energies for the two hydrogen *1s* related states. The lower hydrogen *1s* state has energy located deep in the valence band, approximately 20 eV below the VBM. Thus, it is degenerate with a lower valence band sub-band. The higher has an energy of approximately 7 eV below the VBM and degenerate with the upper valence band subband. Thus, both the states were occupied. Nevertheless, the hydrogen impurity behaves as an electron donor because it breaks the N-Ga bond, which is occupied by two electrons. Therefore, these two electrons were shifted to hydrogen-related states. In addition, H atom brings one electron that has to be donated to the other states; therefore, the H atom can compensate for the Mg acceptor. Therefore, the removal of hydrogen from this state leads to the activation of Mg, at he process invented by Nakamura.

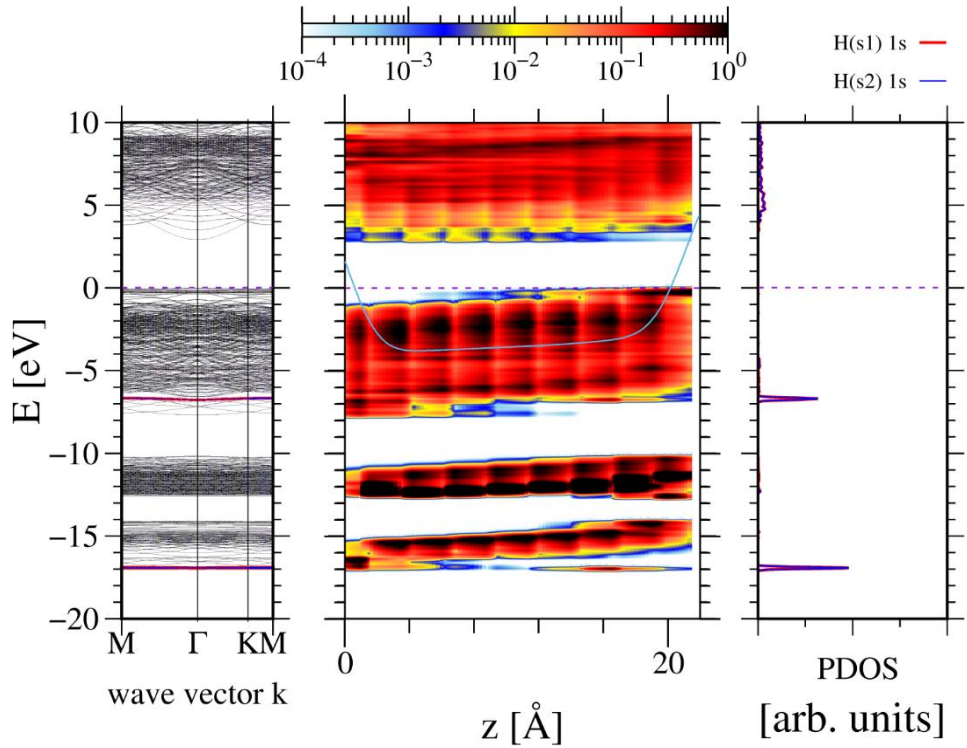

*Fig.4. Band diagram$(2\sqrt{3} \times 2\sqrt{3})$ 8 double Ga-N atom layer slab, obtained in full spin ab initio calculations, representing (0001) GaN surface with attached 11 H atoms ($\theta_H = 11/12$): left – in momentum space, middle – in real space, plotted along c-axis, surface is at the right edge of the diagram, right – projected density of states (PDOS) of two different spin value hydrogen 1s states. The scale of colors at the top represents the density of states in real space.*

The obtained results for the H impurity atom located 2 DALs below the GaN (0001) surface are fully compatible with the results obtained in the H impurity simulation results presented in references 42 and 43 [42,43]. In addition, the electric potential and band profiles shown for both n-type and p-type GaN in Figs 2 and 4, respectively, indicate that the defects are electrically neutral. Therefore, it is not affected by the band-bending subsurface potential caused by Fermi level pinning at the surface and the creation of a subsurface dipole layer [47]. Therefore, this slab can be used to determine the energy path between the bulk and the surface. For these calculations, the properties of hydrogen must be determined, as described in the following subsection.

### b. Hydrogen at GaN(0001) surface

Hydrogen adsorption at the Ga-terminated (0001) $GaN$ surface leads to disintegration of the $H_2$ molecule and attachment of a single H atom in the on-top position. The process was investigated intensively using a large number of experiments and *ab initio* simulations. Experimentally, hydrogen was found to be attached in atomic form, and the adsorption energy was determined [48]. However, the adsorption energy varies, and the dissociative adsorption and molecular desorption barriers are scattered depending on the measurements [48]. For instance, Yang et al. reported an activation barrier for $D_2$ desorption $\Delta E_a^{des} = 2.0 \mp 0.1 \, eV$ [49]. On the other hand, Wampler and Myers determined the full energy diagram of hydrogen in bulk GaN, GaN(0001) surface, and vapor [9]. They have found the energy of molecular hydrogen in the vapor and atomic hydrogen at GaN(0001) surface almost at the same energy level (the difference is mere 0.06 eV/atom), the barrier for this transition between these states equal to $\Delta E_a^{des} = 0.90 \, eV/atom$ and $\Delta E_a^{ads} = 0.96 \, eV/atom$, i.e. $\Delta E_a^{des} = 1.80 \, eV/molecule$ and $\Delta E_a^{ads} = 1.92 \, eV/molecule$. Many other different values were obtained in these measurements, which were precisely and critically discussed in the review by Bermudez [48]. The correct answer is that these scattered data are not due to errors in the experiment, but result from the physics of the process.

Theoretical investigations can be divided into early and late periods [47]. In the early period, hydrogen was found to be adsorbed atomically with fractional coverage $\theta_H = 3/4$ in a particularly stable configuration, satisfying the electron counting rule (ECR) [50,51]. A detailed summary of these studies can be found in Ref 48 [48]. The late period started with the discovery that hydrogen molecule $H_2$ adsorption at $(2 \times 2)$ slab representing $GaN(0001)$ surface is barrierless, dissociative having two different adsorption energies: $\Delta E_a^{ads}(H_2) = 2.2 \, eV/molecule$ and $\Delta E_a^{ads}(H_2) = 0.4 \, eV/molecule$ for 1 initially adsorbed H atom, i.e. for the final coverage $\theta_H = 3/4$ and for 2 initially adsorbed H atoms, i.e. for the final coverage $\theta_H = 1$ [52]. These results were confirmed by atomic adsorption energies showing that adsorption of atomic hydrogen energy is $\Delta E_a^{ads}(H) = 3.4 \, eV/atom$ and $\Delta E_a^{ads}(H) = 1.4 \, eV/atom$ for $\theta_H < 3/4$ and $\theta_H > 3/4$, respectively [52]. Further investigation showed that the hydrogen attached to $GaN(0001)$ surface does not interact, and the transition between these two values is jump-like [45,46]. Extensive investigation has revealed that the adsorption energy difference stems from the difference in the pinning of the Fermi level at the surface [45-47]. According to ECR rule, for $\theta_H \leq 3/4$ Fermi level is pinned by Ga-broken bond state located approximately 0.5 ML below CBM, while for $\theta_H > 3/4$ Fermi level is pinned by H -Ga bonding state located at VBM. The hydrogen atom brings one electron, which is added to 3/4 electrons from the Ga-broken bond state creates 1/4 electron deficiency in the Ga-H bond. In the first case, the missing electron is shifted from the Ga broken bond state, creating an energy gain of approximately 3 eV. In high-hydrogen-coverage Ga, the broken bond state is empty, and no energy gain is possible. This translated into the adsorption energy gain during the adsorption of the H$_2$ molecule, which was finally determined as

i) $\Delta E_a^{ads}(H_2) = 2.24\ eV/molecule$ for $E_F = Ga - broken\ bond$
ii) $\Delta E_a^{ads}(H_2) = 2.62\ eV/molecule$ for $E_F = CBM$
iii) $\Delta E_a^{ads}(H_2) = -2.38\ eV/molecule$ for $E_F = VBM$

Thus, for low hydrogen coverage, the energy gain is positive, whereas for high coverage, it is negative. Thus, the molecular desorption of hydrogen is energetically favorable. The H$_2$ desorption from the surface is the fastest for high H coverage of the surface, i.e. for high hydrogen pressure in the vapor, the opposite of the observed activation of Mg acceptors in annealing in a hydrogen-free ambient. In contrast, annealing in hydrogen vapor leads to the deactivation of acceptors and loss of hole conductivity.

In summary, these results explain numerous problems encountered in the determination of adsorption energy and adsorption/desorption energy barriers [48.49]. The obtained energies differed depending on the microstate of the GaN(0001) surface and the position of the Fermi level. There was no error in the measurements, and the obtained results were physically sound depending on the Fermi level.

The configuration of H attached at GaN surface is presented in Fig. 5.

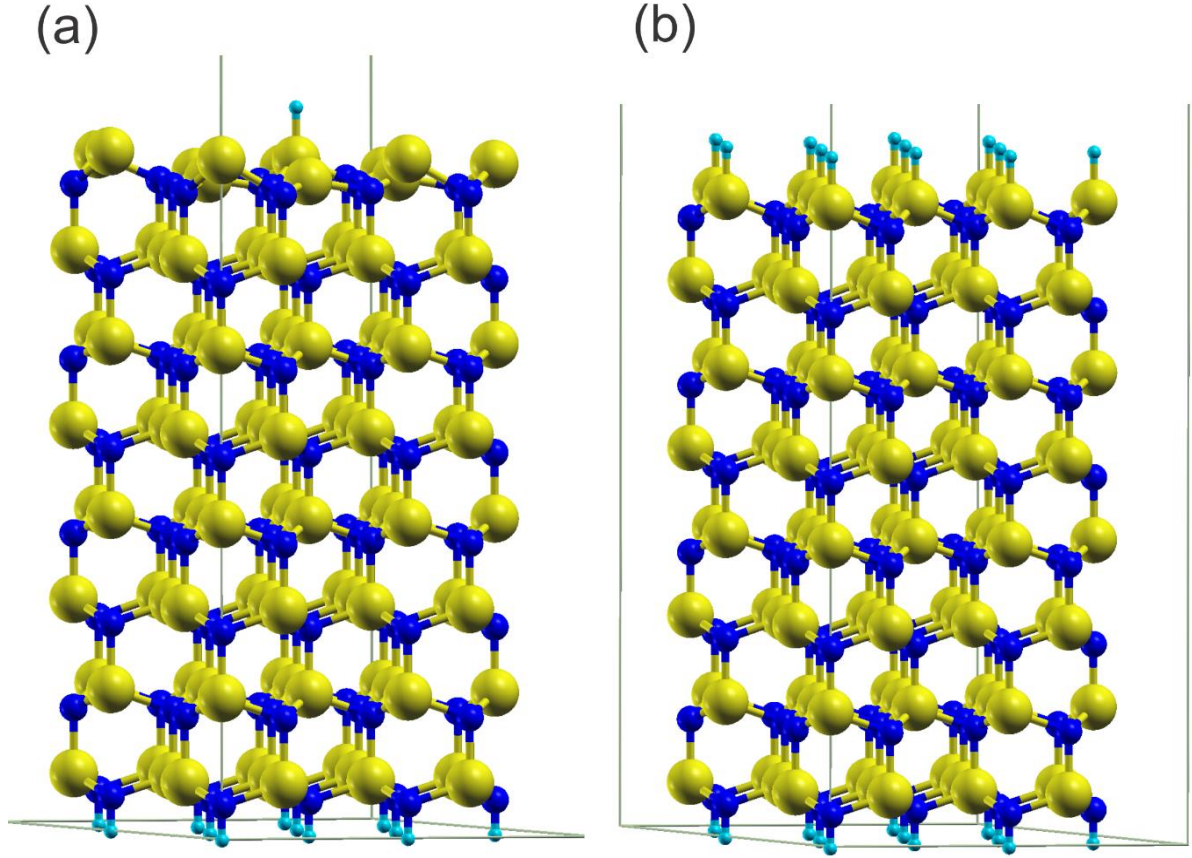

*Fig. 5.Side view of $(2\sqrt{3} \times 2\sqrt{3})$ 8 double Ga-N atom layer slab, representing H atom attached of GaN(0001) surface: (a) originally clean with 1H adatom attached, i.e. effectively $\theta_H = 1/12$, (b) originally covered by 11H adatom with 1H adatom attached, i.e. effectively $\theta_H = 1$. The opposite surface is terminated by fractional charge $(Z = 3/4)$ atoms.*

The simulation results confirm that the attachment of the H adatom to the GaN (0001) surface is identical, irrespective of the additional H coverage. In the case of the originally clean surface, the interatomic distance between the H adatom and the underlying Ga top atom was $d_{Ga-H} = 1.567$ Å while in the case of full coverage, this distance was $d_{Ga-H} = 1.553$ Å. The difference between these two data is $\Delta d_{Ga-H} = 0.014$ Å, relatively small. Moreover, it was shown that the H adatom stabilizes the GaN(0001) surface with a perfect top Ga atom arrangement, similar to the wurtzite lattice. The position of the top Ga atoms in the 1H-covered surface varies considerably because of $sp^3 - sp^2$ reconstruction, which induces atom replacements much larger than the difference in the Ga-H distance [53].

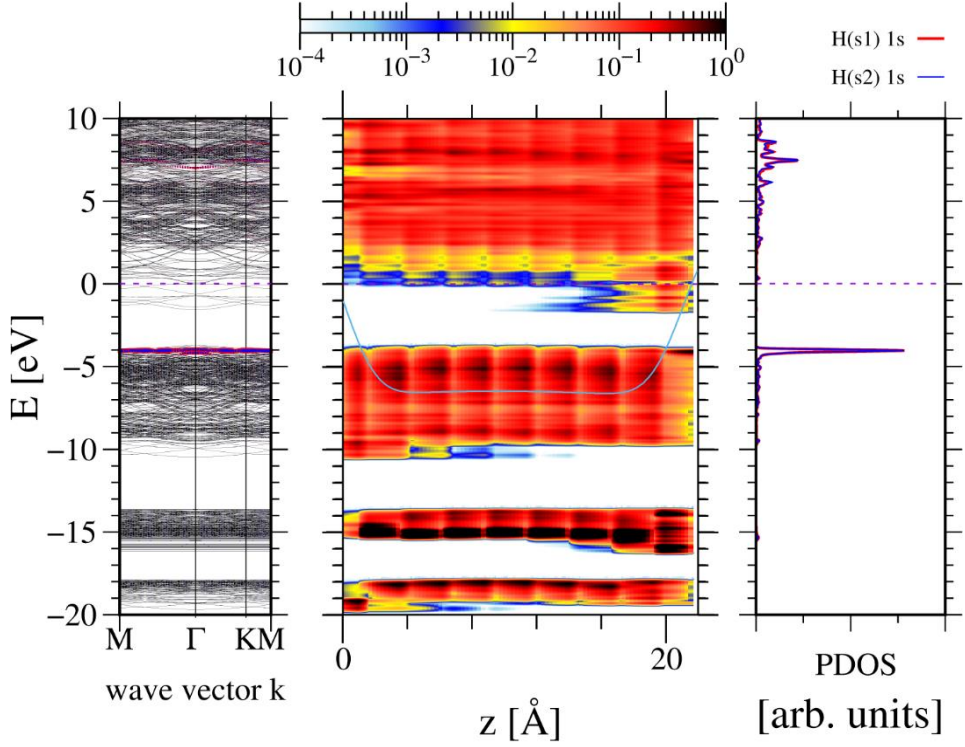

*Fig.6. Band diagram$(2\sqrt{3} \times 2\sqrt{3})$ 8 double Ga-N atom layer slab, obtained in full spin ab initio calculations, representing (0001) GaN surface with attached 1 H atom ($\theta_H = 1/12$): left – in momentum space, middle – in real space, plotted along c-axis, surface is at the right edge of the diagram, right – projected density of states (PDOS) of two different spin value hydrogen 1s states. The scale of colors at the top represents the density of states in real space.*

The electronic properties of the system were determined using these calculations. The band structure of a clean GaN(0001) surface with a single H atom attached is shown in Fig 6. As expected, the Fermi level is pinned to the fractionally occupied Ga broken bond states at the CBM. The double spin Ga-H states were located at the VBM. The system with the H atom located at the surface has a total energy difference of 2.86 eV lower than that in the bulk. Thus, the energy difference is $\Delta E_{s-b}(H) = E_s - E_b = -2.86\ eV$. Thus, the escape of hydrogen from the surface was energetically favorable. Moreover, the energy gain compensates for the pair of hydrogen atoms $\Delta E_{s-b}(2H) = -5.72\ eV$ compensates the energy barrier for desorption which is $\Delta E_a^{ads}(H_2) = 2.24\ eV/molecule$ so the entire process is likely to be very effective, even at relatively low temperatures.

The electronic properties of the fully H-covered GaN(0001) surface are shown in Fig. 7. As it was shown, Fermi level is pinned by Ga-H state at VBM. The Ga broken bond states disappeared. The spin-degenerate states are in the VBM, thus confirming the general assumption of the model.

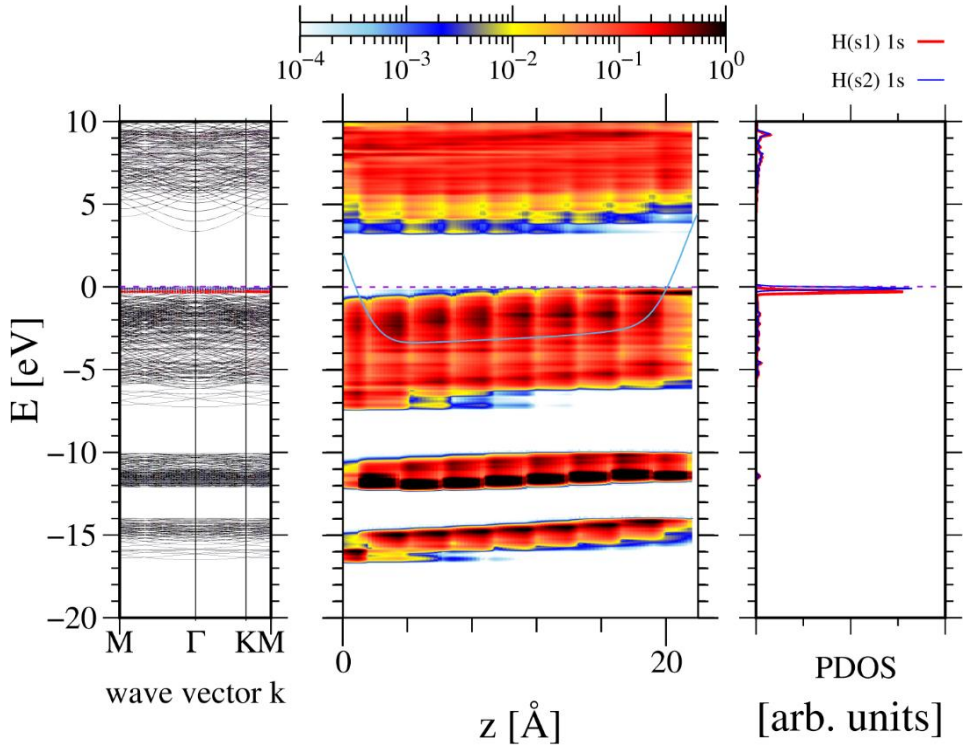

*Fig.7. Band diagram $(2\sqrt{3} \times 2\sqrt{3})$ 8 double Ga-N atom layer slab, obtained in full spin ab initio calculations, representing (0001) GaN surface with attached 12 H atom ($\theta_H = 1/12$): left – in momentum space, middle – in real space, plotted along c-axis, surface is at the right edge of the diagram, right – projected density of states (PDOS) of two different spin value hydrogen 1s states. The scale of colors at the top represents the density of states in real space.*

The system with the atom H located at the surface has its total energy difference 1.11 eV higher than in the bulk. Thus, the energy difference is $\Delta E_{s-b}(H) = E_s - E_b = 1.11\ eV$. Therefore, the escape of hydrogen to the surface is energetically difficult.

An important issue was related to the energy contribution to the energy balance in these two cases. In fact, the H atom in the bulk of n-type GaN (clean GaN (0001) surface) is located in the channel, with its electronic state located in the bandgap. In the case of the H atom in p-type GaN (11 H adatoms attached), the quantum states are located deep in the valence band. In the case of the H atom at the surface, the electron states are identical and located at the VBM for both cases. Thus, the energy difference between the quantum states of the H atom in the bulk is of the order of 10 eV. This is translated into 3.97 eV in favor of clean surface. These states are used in the simulation of the energy profiles along the path from the bulk to the surface using the nudged elastic band (NEB) method [32-34].

### c. Path to the surface – energy barriers

The energy barrier for the bulk-surface transition was determined using the elastic band (NEB) method developed by Henkelman et al. [32-34]. The procedure finds the minimal energy path while keeping the ends of the path fixed and probing the intermediate points for the minimal energy. The necessary condition is to use an energetically stable point at both ends. Consequently, the saddle energy point can be determined. The procedure does not differentiate between the initial and final points, finding the maximal energy point along the transition path, that is, assuring the microscopic reversibility of the transition.

The obtained energy dependencies along the paths are shown in Fig. 8. In these profiles, a common energy scale was set up for the final point of the hydrogen atom attached to the on-top position above the surface Ga atom. This is useful for the comparison of the path, based on the fact that the occupation of Ga-H bonding states is identical for both cases, indicating the same energy of hydrogen for these two cases. This assumption was used to draw the diagram shown in Fig. 8.

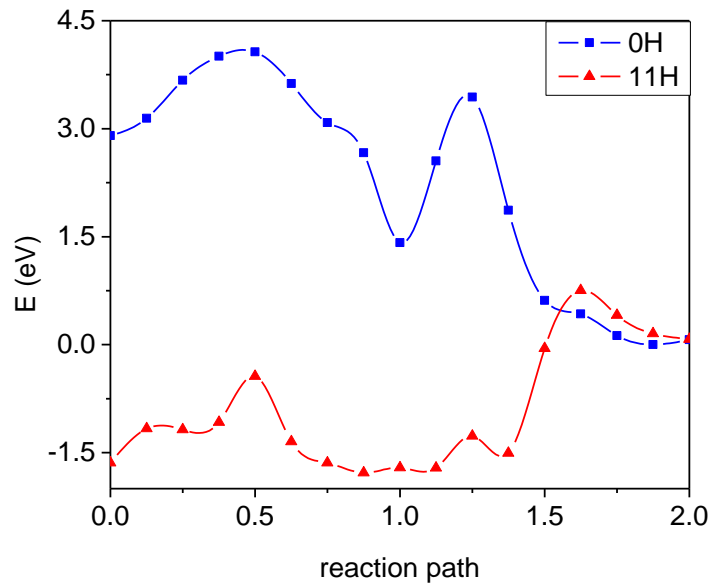

*Fig. 8. Energy profiles obtained for the hydrogen atom outdiffusion path from the bulk to the surface. The final position energy is set common for both cases.*

The obtained energy profiles show a large energy difference at the initial point as a function of the position of the Fermi level at the surface. This is an example of the dependence of the energy barrier of classical diffusion on quantum statistics formulated in Ref. [54, 47]. The presently considered problem corresponds to scenario (i), in which the Fermi level position affects the equilibrium configuration of the diffusive species. The initial point energy for 11H coverage (p-type) was approximately 4.544 eV higher than that for a clean surface. Thus, the bulk energy difference was large, reflecting the difference in the energies of the quantum states discussed above. The energy differences between the surface and bulk are $\Delta E_{s-b}(0H) = -2.835\ eV$ and $\Delta E_{s-b}(11H) = 1.717\ eV$ for 0H and 11H, respectively. Thus, the transition of H atoms from the bulk to the surface is energetically favorable for a clean surface. In the case of an H-covered GaN(0001) surface, the transition requires an energy contribution for the other degrees of freedom. It is noteworthy that the migration profiles were different. For p-type GaN, the barrier inside is almost flat, whereas for n-type GaN, these extrema are more pronounced.

From these simulations, it follows that the control step of Mg activation by hydrogen removal is related to the energy difference observed at a very small distance from the surface. In both cases drastic energy change is observed in the final path fraction of about 0.5 nm, i.e. the creation of surface bond. A bond between Ga and the H adatom was created, and the electrons were transferred. In the case of n-type GaN this finale stage energy difference is about 3.20 eV which could be fractionally used to overcome the surface-vapor molecular desorption energy difference: $\Delta E_a^{ads}(H_2) = 2.24\ eV/molecule$. The initial, deeper portion of the calculated path is essentially energetically neutral, as it is related to the interaction between the H atom and the surrounding GaN lattice. It is different in character, confirming the already determined difference of H diffusion in the bulk: higher and lower barriers for n- and p-type, as has already been identified [42].

### d. Thermodynamics of hydrogen at GaN(0001) surface and in the vapor

The stability of the GaN(0001) surface depends critically on the presence of adsorbates such as hydrogen [55-57] ammonia, or liquid gallium [58]. As shown relatively early, the presence of liquid gallium droplets leads to the dissolution of the GaN microcrystals. In the absence of liquid gallium, these crystals remain stable for a long time and can be observed even by the naked eye [58]. The presence of hydrogen leads to a similar acceleration of GaN decomposition as in the case of liquid Ga. The ammonia overpressure stabilizes GaN polar surfaces almost indefinitely; nevertheless, it could not be used to stabilize the GaN surface during the activation of p-type GaN as it provides an H-rich ambient. Therefore, the annealing process must be investigated thermodynamically only in the hydrogen context.

In the past, the adsorption of molecular hydrogen on the GaN(0001) surface was intensively investigated. The adsorption energy of the species was defined as the difference between the total *ab intio* energies in the vapor and at the surface: $\Delta E_{DFT}^{ads} \equiv h^v - h^s$. In addition, it has been shown that molecular hydrogen is adsorbed dissociatively in the atomic hydrogen on-top positions ($H_2(g) \rightarrow 2H_{s-OT}$) with the following energy gain [45,46].:

- For $E_F = E(Ga_{BB})$

$$\Delta E_{DFT-Ga_{BB}}^{ads}(H_2) = 2E_{DFT}(H^s) = 2.314\ eV \qquad (1a)$$

- For $E_F = VBM$

$$\Delta E_{DFT-VBM}^{ads}(H_2) = 2E_{DFT}(H^s) = -2.286\ eV \qquad (1b)$$

where $Ga_{BB}$ denotes the Fermi level pinned at the broken Ga bond. The latter is related to the creation of a Ga-H bond state with energy close to the valence band maximum (VBM). As argued above, electrons are shifted from Ga broken bond states to Ga-H states so that the electron energy gain is close to the bandgap, i.e. $E_g^{GaN} = 3.47\ eV$. This electron transfer is exhausted at $\theta_H = 0.75\ ML$ therefore, for higher coverage, the energy gain is negative, i.e. no attachment is possible.

The adsorption of hydrogen was simulated using *ab initio* data on the energy of adsorption and vibrational properties of the H adatom [60, 61]. This adds a zero-point energy contribution of $\Delta E^{ZPE}(H) = 0.27812\ eV$. The temperature-dependent contribution to the chemical potential of H can be obtained from the summation over the vibrational spectra using the following relation:

$$F_{H^s}^{vib}(T) = k_B T \sum_j \ln[1 - exp(x_j)] \qquad (2)$$

where $x_j \equiv \frac{\hbar \omega_j}{k_B T}$ and $\omega_j$ is the phonon frequency of the j-th phonon mode. The resulting temperature-dependent free energy of the H adatom can be approximated by the following relation [59,60]:

$$F_{H^s}^{vib}(y) = 0.27812 + 0.0068y - 0.128y^2 + 8.66 \times 10^{-3}y^3 \qquad (3)$$

where $y \equiv T/1000$ and the free energy and temperature are in eV and Kelvin, respectively. The data indicate a shallow maximum at low temperatures, related to the stiffening of the lattice due to H adsorption. These data may be used to obtain the chemical potential of the H adatom as a function of temperature and coverage as follows:

$$\mu_H(y, T, \theta_H) \cong [\Delta E_{DFT}^{ads}(2H)]/2 + F_{H^s}^{vib}(y) + k_B T \ln\left(\frac{\theta_H}{1-\theta_H}\right) \qquad (4)$$

To determine the pressure coverage diagram, these data may be compared with the above-derived values for the hydrogen chemical potential in the vapor:

$$\mu_H(y, T, \theta_H) = \frac{1}{2} \mu^v_{H_2}(p_{H_2}, y, T) \quad (5)$$

where the pressure values obtained above are used to obtain the coverage pressure dependence for selected temperatures, relevant for MOVPE growth of nitrides via the equation:

$$k_B T \ln\left(\frac{\theta_H}{1-\theta_H}\right) = -\left[\Delta E^{ads}_{DFT}(H_2)\right]/2 - F^{vib}_{H^s}(y) + \frac{1}{2} F^v_{H_2}(y) + \frac{k_B T}{2} \ln(p_{H_2}) \quad (6)$$

The important issue is the determination of the boundaries of the region at which the Fermi level undergoes a jump owing to the adsorption energy jump given in Eq. 1, which occurred at a critical coverage $\theta_H = 3/4$. These pressures were determined using the following relations (Eq. 1.):

- For $E_F = E(Ga_{BB})$

$$p_{H_2-GaBB}(y) = 3 \exp\left\{\left[\Delta E^{ads}_{DFT-Ga}(H_2) + 2F^{vib}_{H^s}(y) - F^v_{H_2}(y)\right]/kT\right\} \quad (7a)$$

- For $E_F = VBM$

$$p_{H_2-VBM}(y) = 3 \exp\left\{\left[\Delta E^{ads}_{DFT-VBM}(H_2) + 2F^{vib}_{H^s}(y) - F^v_{H_2}(y)\right]/kT\right\} \quad (7b)$$

These data are presented in Fig. 8 for a wide temperature range, showing that the region between these two lines, corresponding to the Fermi level free, extends over most of the experimentally used thermodynamic conditions.

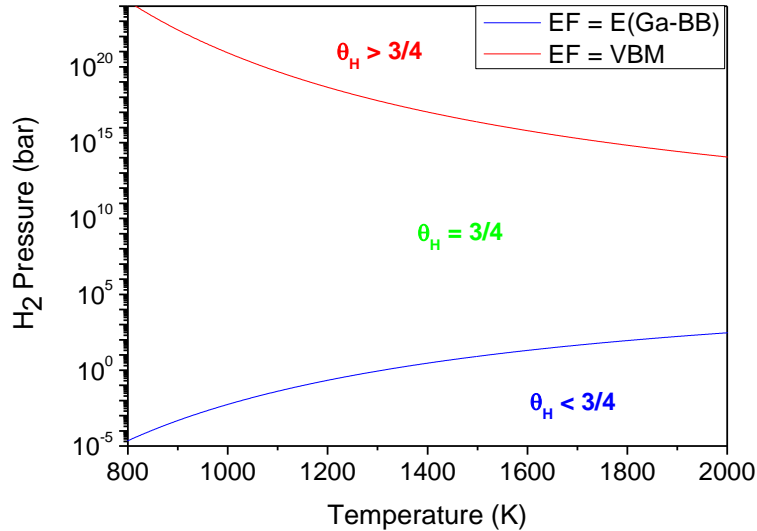

Fig. 9. Pressure-temperature dependence of the critical hydrogen coverage ($\theta^{cr}_H = 3/4$) of GaN(0001) surface.

From these data, it follows that the pressure range corresponding to the critical coverage $\theta_H^{cr} = 3/4$, and accordingly, the Fermi level is free, extends over the entire range of conditions relevant to MVPE, HVPE, and p-type activation. The borders of these regions were determined as follows.

- For $E_F = E(Ga_{BB})$

$$p_{H_2-GaBB}(y) = 3\,exp\left\{\left[-1.9032 + 1.366y + 4 \times 10^{-3}y^2 - 6.68 \times 10^{-3}y^3\right]/0.0862y\right\} \quad (8a)$$

- For $E_F = VBM$

$$p_{H_2-VBM}(y) = 3\,exp\left\{\left[2.6938 + 1.366y + 4 \times 10^{-3}y^2 - 6.68 \times 10^{-3}y^3\right]/0.0862y\right\} \quad (8b)$$

where $y \equiv T/1000$ and the temperature is in Kelvin. It is worth noting that the pressures corresponding to the coverage in the ranges $\theta_H < 0.75$ and $\theta_H > 0.75$ are separated by large pressure interval, as shown in Fig. 9.

The critical pressure is the higher molecular hydrogen pressure at which the Fermi level is still pinned in the Ga-broken bond state (i.e. blue line in Fig. 9). It is worth to note that the coverage $\theta_H \cong 0.75$ for $E_F = E(Ga_{BB})$ is attained for the hydrogen pressures $p(H_2) = 0.0408\,bar$ for $T = 1100\,K$ and $p(H_2) = 0.87\,bar$ for $T = 1300\,K$. The $T = 1100\,K$ it typical for the annealing of Mg-doped acceptors to activate p-type material. Based on this, the maximum hydrogen pressure for effective p-type annealing can be determined as

$$p_{H_2}^{max}(T) = \exp(-77.863 + 0.1519T - 1.079 \times 10^{-4}T^2 + 2.862 \times 10^{-8}T^3) \quad (9)$$

where the pressure and temperature are in the bars and Kelvins, respectively. The temperature range extends from 800 K to 1100 K, and the pressures are $p_{H_2}^{max}(800K) = 2.22 \times 10^{-5}\,bar$ and $p_{H_2}^{max}(1100K) = 0.04\,bar$. These dependencies may serve as thermodynamic conditions for successful annealing processes of Mg activation, as invented by Nakamura. The functional dependence given in Eq. (9) provides a reasonable approximation that can be used for any temperature. Thus, the thermodynamic conditions, that is, i.e. temperature and molecular hydrogen pressure for Nakamura, were fully determined.

An additional issue, which is less directly related to annealing but still relevant, is the surface H-coverage corresponding to the Fermi level pinned at the VBM and also for the Ga-broken bond state. The H coverage for the two selected temperatures, $T = 1100\,K\,(y = 1.1)$ and $T = 1300\,K\,(y = 1.3)$, obtained from Eq. 8 are:

- For $E_F = E(Ga_{BB})$

$$p_{H_2-GaBB}(\theta_H, y) =$$
$$\left(\frac{\theta_H}{1-\theta_H}\right)^2 exp\left\{\left[-1.9032 + 1.366y + 4\times10^{-3}y^2 - 6.68\times10^{-3}y^3\right]/0.0862y\right\} \quad (10a)$$

- For $E_F = VBM$

$$p_{H_2-VBM}(\theta_H, y) =$$
$$\left(\frac{\theta_H}{1-\theta_H}\right)^2 exp\left\{\left[2.6938 + 1.366y + 4\times10^{-3}y^2 - 6.68\times10^{-3}y^3\right]/0.0862y\right\} \quad (10b)$$

The data obtained from these dependence are presented in Fig. 10.

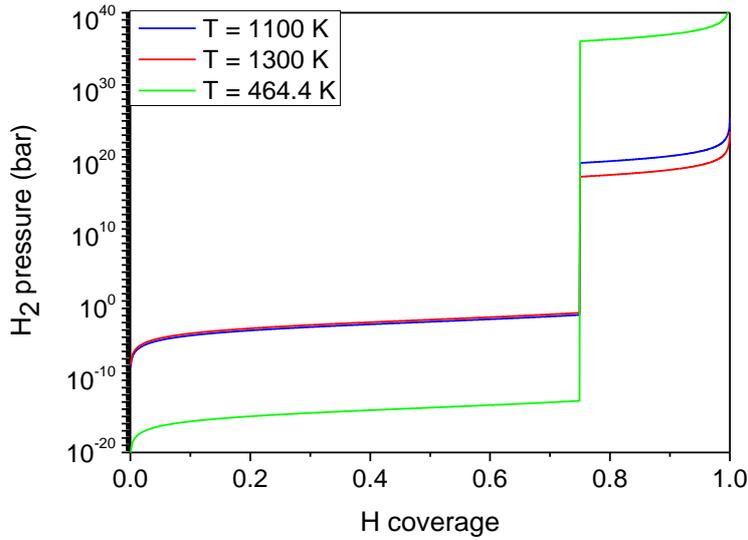

*Fig. 10. Hydrogen pressure-coverage dependence at GaN(0001) surface.*

These data may serve as guidance for the further determination of Gan epitaxy by MOVPE and HVPE methods, especially in the case of hydrogen transport gas.

## IV. Summary and conclusions.

*Ab intio* based thermodynamic analysis of GaN:H (bulk)-GaN(0001)-H surface-H$_2$(vapor) composite system was performed to obtain a microscopic picture and the thermodynamic conditions corresponding to Mg annealing activation of p-type GaN of MOVPE-grown GaN. The results provide a detailed picture of the system, allowing the determination of the conditions for processing GaN to obtain the hole conductivity. The summary and importance of this research are the best understood by the presentation of the state of the art **before** publication, the results obtained **in**, and the state of the art **after** publication [53,54, 47].

The state of the art **before,** in the area related to the publication:

(i) The thermodynamic procedure of Mg acceptor activation was reliably determined and can be summarized as a high-temperature hydrogen-free process,

(ii) Mg activation was determined to be caused by the removal of hydrogen from GaN to the vapor by a second-order process.

(iii) Hydrogen impurities in GaN crystals were found to be dependent on the Fermi level, with two states clearly identified.

(iv) The first activation stage, that is, thermal mobility induced by a high activation temperature, was identified

(v) Several hypotheses regarding the microscopic mechanism of hydrogen removal are considered plausible candidates for the activation scenario.

(vi) No microscopic picture of the annealing process was formulated.

The results obtained **in** the publication :

(i) It was shown that hydrogen escape is controlled by the pinning of the Fermi level at the surface; at CBM, escape is possible, and at VBM, it is blocked.

(ii) Confirmed critical H coverage of Fermi level pinning: for $\theta_H < 3/4$ - $Ga_{BB}$ close to the CBM, for $\theta_H > 3/4$ - VBM,

(iii) The energy migration profile across the two topmost Ga-N atomic layers was determined, showing its control of H escape from GaN.

(iv) H impurities are neutral; therefore, the surface electric potential and band bending do not affect the escape of H from the GaN.

(v) The energy barrier for H escape emerges in the Ga-H bond creation stage

(vi) The molecular desorption of $H_2$ from the surface is not the rate-determining step of p-type activation.

(vii) hydrogen maximal partial pressure in the vapor as a function of temperature for the activation of p-type GaN was obtained.

(viii) The first dependence of the diffusion process on the Fermi level position is determined theoretically and confirmed experimentally.

The state of the art **after,** in the area related to the publication:

(i) The detailed description of Nakamura Mg activation process was formulated,

- (ii) The process was controlled by pinning the Fermi level at the fractionally H-covered GaN(0001) surface.
- (iii) The equilibrium state of $GaN(0001) - H_2$ system was fully determined, allowing the determination of p-type GaN activation molecular hydrogen partial pressure–temperature conditions.

In conclusion, it could be stated that Nakamura Mg activation is elucidated, the thermodynamic conditions for the process were determined thus the state of the art is considerably progressed proving that the present work contributes to the overall knowledge in this area significantly.

**Acknowledgement**

The calculations reported in this paper were performed using the computing facilities of the Interdisciplinary Centre for Mathematical and Computational Modelling of Warsaw University (ICM UW) under Grant GB84-23. We gratefully acknowledge the Polish high-performance computing infrastructure PLGrid (HPC Centers: ACK Cyfronet AGH) for providing computer facilities and support within the computational grant no. PLG/2024/017466.